\documentclass[aps,pra,preprint,groupedaddress]{revtex4-1}

\begin{document}
\newcommand{\comm}[1]{\underline{\tt #1}}

\newcommand{\modegy}{\,({\rm mod}\,1)}
\newcommand{\atn}{\tan^{-1}}
\newcommand{\ch}{{\rm ch}}
\newcommand{\sh}{{\rm sh}}
\newcommand{\bml}{\begin{mathletters}}
\newcommand{\eml}{\end{mathletters}}
\newcommand{\be}{\begin{equation}}
\newcommand{\ee}{\end{equation}}
\newcommand{\ba}{\begin{array}}
\newcommand{\ea}{\end{array}}
\newcommand{\bea}{\begin{eqnarray}}
\newcommand{\eea}{\end{eqnarray}}
\newcommand{\nn}{\nonumber}
\newcommand{\smi}{\!-\!}
\newcommand{\spl}{\!+\!}
\newcommand{\pr}{\prime}
\newcommand{\dpr}{\prime\prime}
\newcommand{\cS}{{\Delta\cal S}}
\newcommand{\OL}{\overline}
\newcommand{\cK}{{\cal K}}
\newcommand{\cL}{{\cal L}}

\def\eqalign#1{
\null \,\vcenter {\openup \jot \ialign {\strut \hfil $\displaystyle {
##}$&$\displaystyle {{}##}$\hfil \crcr #1\crcr }}\,}

\title{
On the normalization of the partition function \\
of Bethe Ansatz systems}
\author{\large F.~Woynarovich\footnote{e-mail: fw@szfki.hu}}
\affiliation{Institute for Solid State Physics and Optics\\
Hungarian Academy of Sciences\\
1525 Budapest 114, Pf 49.}

\begin{abstract}
In this note I revisit the calculation of partition function of simple one-dimensional systems solvable by Bethe Ansatz. Particularly I show that by the precise definition and treatment of the partition function the nontrivial normalization factor proposed in a recent work to give the correct $O(1)$ corrections to the free energy can be derived in a straightforward manner. 
\end{abstract}


\maketitle

\setlength{\parskip}{2ex}
\setlength{\parindent}{2.5em}
\setlength{\baselineskip}{3ex}

\vfill\eject
\section{Introduction}
\label{sec:Int}

Recent developments in both field theory and solid state physics have shown, that in certain problems, where surface or impurity effects are important, non-macroscopic contributions to the thermodynamic quantities like the free energy may play an important role \cite{AffLud,Kondo}.
Among these systems those solvable by Bethe Ansatz (BA) due to the exact treatability are of special importance. For these one-dimensional (1D) systems the free energy is calculated following the method developed by C.N.~Yang and C.P.~Yang \cite{CNYCPY} 
for the $\delta$ Bose gas. The basic idea of this method is that through the density of momenta (rapidities) an entropy can be defined and a free energy as a functional of the rapidity density can be constructed. The minimization of this functional yields both the equilibrium state and the macroscopic part of the free energy of the finite temperature system. Based on this idea the present author developed a method to calculate the $O(1)$ free energy contributions of the states near to the equilibrium \cite{Woyna}, however the contributions found have not met the expectations \cite{DoFiChTa}: for periodic boundary conditions (PBC) no $O(1)$ corrections have been expected, but the calculation gave some, and also for the case of open ends with integrable boundaries (IB) some of the obtained terms were not of the expected structure. Recently, based on intuitive arguments concerning the density of states in the configuration space a nontrivial normalization of the partition function has been proposed \cite{Pozsgay}, by which these differences can be dissolved: in the case of PBC the corrections are canceled, while for the IB case they are corrected. In the present note we derive this nontrivial factor directly by the careful definition and treatment of the partition function. This way this work confirms the proposal of \cite{Pozsgay} and completes \cite{Woyna}.

As a starting point we briefly review the ingredients of the calculations. This will serve also to make clear our notations and expose the problem in a more tractable form. It has been tempting to formulate our treatment in a general form, as however the derivation of the entropy term for the free energy functional is slightly different for closed and open boundaries we treat the two cases separately:
first we deal with the case of periodic boundary condition in more details, and in case of integrable boundaries we point out the differences only.

\section{BA and BA thermodynamics}
\label{sec:BApf}

We consider a system with BA equations 
\be\label{BAe}
Lp(\theta_i)+\sum_{j=1}^N\phi(\theta_i-\theta_j)=2\pi I_i.
\ee
Here the $\theta$ are the rapidities of the particles, $p(\theta)$ is the momentum of a particle with rapidity $\theta$, the   
$\phi(\theta-\theta')$ is a phase closely related (up to a constant equal) to the phase shift arising when a particle with rapidity $\theta$ is scattered on an other with $\theta'$, and the $I$ quantum numbers are either integers or halfs of odd-integers depending on the number of particles $N$. (For later purposes we chose the Riemann sheets to have $\phi$ continuous at zero argument.) To each set of quantum numbers $\{I_j, I_j\not= I_l,\ \mbox{if}\ j\not= l\}$ (\ref{BAe}) defines a set of real rapidities $\{\theta_j, \theta_j\not= \theta_l,\ \mbox{if}\ j\not= l\}$. The wave functions belonging to the different solutions of (\ref{BAe}) are orthogonal and form a complete set.
The energy of such a state (modified due to the chemical potential $\mu$ if necessary) is the sum of the contributions of the individual particles
\be\label{energy}
E-\mu N=E(\theta_1,\theta_2,\dots,\theta_N)= \sum_{i=1}^N {e}(\theta_i).
\ee

The finite temperature description \cite{CNYCPY} is based on the idea, that for a macroscopic system the roots of (\ref{BAe}) can be described by densities, and the thermodynamic quantities can be given by these. In particular the $\rho(\theta)$ density of particles is defined so, that the 
$\#\ \mbox{of}\ \theta_i\in(\theta,\theta+\Delta\theta)\ \mbox{is}\ L\rho(\theta)\Delta\theta$ and the $\rho_h(\theta)$ density of holes is given by the equation obtained from (\ref{BAe})
\be\label{densities} 
\rho(\theta)+\rho_h(\theta)=\frac{1}{2\pi}\frac{\partial p(\theta)}{\partial\theta}+\int K(\theta-\theta')\rho(\theta')d\theta'\quad\mbox{with}\quad K(\theta)=\frac{1}{2\pi}\frac{\partial \phi(\theta)}{\partial\theta}.
\ee
All $\rho(\theta)$ functions are physical, for which (\ref{densities}) yields a non-negative $\rho_h(\theta)$, as for these densities $I_i$ quantum number sets can be constructed which define $\theta_i$ roots distributed according to $\rho(\theta)$ (with a certain accuracy). The number how many ways this can be done, i.e.~the number of states represented by one single rapidity density is estimated by the combinatorial factor 
\be\label{Omegarho} 
\Omega[\rho(\theta)]=\prod{L(\rho(\theta)+\rho_h(\theta))\Delta\theta\choose L\rho(\theta)\Delta\theta}.
\ee
Through this the entropy is
\be 
S[\rho(\theta)]=\ln \Omega[\rho(\theta)],
\ee
and a free energy functional can be defined:
\be\label{F[rho]}
F[\rho(\theta)]=E[\rho(\theta)]-TS[\rho(\theta)] \quad \mbox{with}\quad E[\rho(\theta)]=\sum e(\theta)\rho(\theta)\Delta\theta,
\ee
and $T$ being the temperature.
The macroscopic part of this is
\bea\label{F_L}
&&F_L[\rho(\theta)]=
L\int\!{e}(\theta)\rho(\theta)d\theta-\nn\\
&&L\int T\left[(\rho(\theta)\!+\!\rho_h(\theta))\ln(\rho(\theta)\!+\!\rho_h(\theta))\!
-\!\rho(\theta)\ln\rho(\theta)\!
-\!\rho_h(\theta)\ln\rho_h(\theta)
\right]d\theta.
\eea   
The equilibrium density $\rho_0(\theta)$ is given by the minimization of (\ref{F_L}) under the constraint of (\ref{densities}). This gives that in addition to (\ref{densities}) the densities satisfy   
\be\label{densityratio}
\frac{\rho_0(\theta)}{\rho_{h,0}(\theta)}=e^{\displaystyle{-\beta\epsilon(\theta)}}
\ee
too, where $\beta=1/T$, and
\be
\epsilon(\theta)=e(\theta)-T\int K(\theta-\theta')\ln\left(1+e^{\displaystyle{-\beta\epsilon(\theta')}}\right)d\theta'.
\ee
Finally the macroscopic free energy is the minimal value of $F_L$:
\be\label{Fmin}
F_{min}=-\frac{L}{\beta2\pi}\int \frac{\partial p(\theta)}{\partial\theta}\ln\left(1+e^{\displaystyle{-\beta\epsilon(\theta)}}\right)d\theta.
\ee

In \cite{Woyna} we calculated corrections to this macroscopic free energy by calculating the sum
\be\label{Fsum}
\sum_{\rm{all}\ \rho(\theta)}e^{\displaystyle{-\beta F[\rho(\theta)]}}
\ee
with an accuracy enough to have the next to leading order corrections too. This led to 
\be\label{Fsum_result}
\sum_{\rm{all}\ \rho(\theta)}e^{\displaystyle{-\beta F[\rho(\theta)]}}=e^{\displaystyle{-\beta F_{min}+\Delta S}},
\ee
but according to Pozsgay's proposal \cite{Pozsgay} the correct partition function is of the form
\be\label{ZbyPozsgay}
Z={\cal N}\sum_{\rm{all}\ \rho(\theta)}e^{\displaystyle{-\beta F[\rho(\theta)]}}\quad\mbox{leading to}\quad Z={\cal N}e^{\displaystyle{-\beta F_{min}+\Delta S}}
\ee
with ${\cal N}$ being a well defined $O(1)$ factor to be discussed later. The important claim in (\ref{ZbyPozsgay}) is, that (\ref{Fsum}) itself reproduces the partition function up to a normalization factor only. To clarify this we recalculate the partition function starting from the very beginning, as we think, that in a correct treatment all normalizations are defined uniquely in a "natural" way. 

\section{Partition function at periodic boundary conditions}
\label{sec:PB}

The grand canonical partition function is 
\be\label{Z}
Z=\sum_{N=1}^\infty Z_N
\ee
with
\be\label{Z_N}
Z_N=\sum_{I_1<I_2<\dots<I_N}\exp{\left\{-\beta E(\theta_1,\theta_2,\dots,\theta_N)\right\}},
\ee
and our aim is to transform (\ref{Z}-\ref{Z_N}) into a functional of the rapidity densities used in the thermodynamical description. The steps of this are the following.
\begin{enumerate}
	\item First we write the sum in (\ref{Z_N}) in a form, which can be well approximated by an integral. The problem in this is that the requirement $I_i<I_j$ is essentially different for the cases with the $I_j$'s being discrete or continuous variables (see Appendix \ref{sec:summation}), thus one has to represent the $\sum_{I_1<I_2<\dots<I_N}$ in a form, which does not contain such restrictions. This is possible if, as in our case, the summand is symmetric in the variables. The idea is, that from an unrestricted sum we subtract the contributions of the nonphysical ($I_i=I_j$ type) configurations.
The technical details of the procedure are given in Appendix \ref{sec:summation}, here we give the result only:
	\be\label{ossz}
	 \sum_{I_1<I_2<\dots<I_N}f(I_1,I_2,\dots I_N)=\sum_P a(P)\sum_{I_1,\dots,I_n}f^{(P)}(I_1,I_2,\dots I_n).
	\ee
	Here the summation on the right-hand side goes over the partitions $P$ of the number $N$, and in $f^{(P)}$ groups of the $I_j$ parameters corresponding to the elements in $P$ are equal: for example if a partition is $P:\{p_1+p_2+\dots{}+p_n\}$, then 	
  \be\label{ossz2}
  f^{(P)}(I_1,I_2,\dots I_n)=f(\overbrace{\underbrace{I_1,\dots, I_1}_{p_1},\underbrace{I_2,\dots,I_2}_{p_2},\dots,\underbrace{I_n,\dots,I_n}_{p_n}}^N).
  \ee
In the partition $P:\{1+1+\dots+1=N\times1\}$ the summation is taken without restriction over the $I$'s, and the coefficient is $a(P)=\frac{1}{N!}$. For the partition $p:\{2+1+\dots+1=2+(N-2)\times1\}$ in the summation the first two variables of $f$ are kept equal, and the coefficient is $a(P)=\frac{1}{N!}{N\choose2}$. A systematic way to construct the $a(P)$ coefficients and a proposal for their form is given in Appendix \ref{sec:summation}.

The summand $f$ depend on the $I_j$ quantum numbers through the $\theta_i$ variables determined by (\ref{BAe}). For this we have to define the (nonphysical) solutions of (\ref{BAe}) for $I$ sets containing equal quantum numbers too. In order to do this we consider (2.1) as an analytic relation between the elements of the sets $\{I_j\}$ and $\{\theta_i\}$, and for equal $I_j$'s we shall take the solutions for $\theta$'s obtained through the limiting process in which the corresponding $I$'s tend to the required value. In practice this means, that equal $I$'s will define equal $\theta$'s, and each $\theta_j$ will be summed according to its multiplicity, i.e.~for a partition $P:\{p_1+p_2+\dots{}+p_n\}$
\be\label{mBAe}
Lp(\theta_i)+\sum_{j}^n\phi(\theta_i-\theta_j)p_j=2\pi I_i.
\ee
	
	\item Next we replace the summations by integrals:
	\be\label{int_repr}
	\sum_{I_1}\sum_{I_2}\dots\sum_{I_n}\Longrightarrow\int\int\dots\int\prod_{i=1}^n dI_i.
  \ee
For this from now on we have to consider the $I_i$ quantum numbers in (\ref{BAe}) and (\ref{mBAe}) as continuous variables. 

As in our derivation this integral representation of the sums over the quantum numbers has a central role, we have to note on the 
error introduced this way. It seems plausible, that this step is accurate enough, but to give a quantitative estimate on the error needs different considerations for the different models. In Appendix \ref{sec:integrationerror} we analyze the case of the $\delta$ Bose gas with PBC in more details. 

  \item The integration variables are changed from the $I$'s to the $\theta$'s
  \be
	\int\int\dots\int\prod_{i=1}^n dI_i\Longrightarrow\int\int\dots\int\det\left\{\frac{\partial I_i}{\partial\theta_j}\right\}\prod_{j=1}^n d\theta_j
  \ee
The Jacobi determinant is obtained from (\ref{mBAe}):
   \be\label{Jacobi}
   \det\left\{\frac{\partial I_i}{\partial\theta_j}\right\}=\det\{L\varrho(\theta_i)\delta_{ij}-K(\theta_i-\theta_j)p_j\}
   \ee
where
\be
\varrho(\theta_i)=\frac{1}{2\pi}\frac{\partial p(\theta_i)}{\partial\theta_i}+\frac{1}{L}\sum_{j=1}^nK(\theta_i-\theta_j)p_j
\ee
(with $K(\theta)$ given in (\ref{densities})). Factorizing (\ref{Jacobi}) as in \cite{Pozsgay} one has
\be\label{factorPBC}
\det\left\{\frac{\partial I_i}{\partial\theta_j}\right\}=\left(\prod_{j=1}^n L\varrho(\theta_j)\right)D^{(P)}(\theta_1,\dots\theta_n)
\ee
with
\bea         
&&D^{(P)}(\theta_1,\dots\theta_n)=\det\left\{\delta_{ij}-\frac{K(\theta_i-\theta_j)p_j}{L\varrho(\theta_i)}\right\}\equiv\nn\\
&&\det\left\{\delta_{ij}-\frac{K(\theta_i-\theta_j)p_j}{L\varrho(\theta_j)}\right\}.
\eea
It is important to note, that both $\varrho(\theta_j)$ and $\det\left\{\delta_{ij}\!-\!{K(\theta_i\!-\!\theta_j)p_j}/{L\varrho(\theta_i)}\!\right\}$ are continuous in the $\theta$'s. (Although the dimensions of the matrices $\left\{\delta_{ij}-{K(\theta_i-\theta_j)p_j}/{L\varrho(\theta_i)}\right\}$ are different for the cases $\theta_k=\theta_l$ and $\theta_k\to\theta_l$, the corresponding determinants are equal.) 
Due to this, if we denote by $D(\theta_1,\theta_2,\dots,\theta_N)$ the determinant belonging to the case of all $I_j$ (i.e.~all $\theta_j$) different, then $D^{(P)}(\theta_1,\dots\theta_n)$ can be obtained by taking groups of the variables equal
like in (\ref{ossz2}). This way we have  
\bea\label{weightofdtheta}
&&\sum_{I_1<I_2<\dots<I_N}f(I_1,\dots I_N)=\nn\\&&\sum_P a(P){\int\!\dots\!\int} f^{(P)}(\theta_1,\dots \theta_{n_P})D^{(P)}(\theta_1,\dots\theta_{n_P})\!\left(\prod_{j=1}^{n_P} L\varrho(\theta_j)d\theta_j\right)\!,
\eea
where the $a(P)$ coefficients are the same as in (\ref{ossz}).

\item We divide the $\theta$ axis into $\Delta\theta_\alpha$ intervals. If this intervals are small enough (as required according to \cite{Woyna}), all $f(\theta_1,\dots,\theta_N)$, $D(\theta_1,\dots,\theta_N)$ and $\varrho(\theta)$ can be given accurate enough by giving the $N_\alpha$ numbers of $\theta_i$'s  falling into the interval $\Delta\theta_\alpha$ ($N_\alpha=\#\ \mbox{of}\ \theta_i\in\Delta\theta_\alpha$). 
For this we may write 
\be
\varrho(\theta)=\frac{1}{2\pi}\frac{\partial p(\theta)}{\partial\theta}+\frac{1}{L}\sum_{\alpha}K(\theta-\theta_\alpha)N_\alpha,
\ee
where $\theta_\alpha$ is a mean value in $\Delta\theta_\alpha$, and
\bea
&&f(\theta_1,\dots,\theta_N)\longrightarrow f(\{N_\alpha\})
=\exp\left\{-\beta E(\{N_\alpha\})\right\},\\
&&\nn\\
&&E(\{N_\alpha\})=\sum_{\alpha}{e}(\theta_\alpha)N_\alpha\ ,\\
&&D(\theta_1,\dots,\theta_N)\longrightarrow D(\{N_\alpha\})
=\det\left\{\delta_{\alpha\beta}-\frac{K(\theta_\alpha-\theta_\beta)N_\beta}{L\varrho(\theta_\beta)}\right\}\!\!.
\eea
We note here, that for $N_\alpha$ each $\theta_i$ in $\Delta\theta_\alpha$ have to be taken with the multiplicity $p_i$, thus $\sum_\alpha N_\alpha=N$. We note also, that as all variables within a single interval $\Delta\theta_\alpha$ are taken equal to the same mean value $\theta_\alpha$, distinguishing $f(\{N_\alpha\})$ ($D(\{N_\alpha\})$) with respect to $P$ has no meaning any more. 

\item The $\theta$ integrals are completed under the restriction, that the number of $\theta$'s in $\Delta\theta_\alpha$ are $N_\alpha$. Approximating $\varrho(\theta_j)$ in $\left(\prod_{j=1}L\varrho(\theta_j)d\theta_j\right)$ by the same constant $\varrho(\theta_\alpha)$ if $\theta_j\in \Delta\theta_\alpha$, and following the arguments of Appendix \ref{sec:integral} leading to (\ref{integralbyintervals}) (taking also into account, that now due to 
(\ref{weightofdtheta}) each $d\theta_i$ belonging to the integral over $\Delta\theta_\alpha$ is multiplied by $L\varrho(\theta_\alpha)$) the result of the integrals over the intervals  $\Delta\theta_\alpha$ is 
\be\label{thetaintegralresult}
\prod_\alpha{L\varrho(\theta_\alpha)\Delta\theta_\alpha\choose N_\alpha},
\ee
thus
\be
Z_N=\sum_{\left\{\sum N_\alpha=N\right\}}
\exp\left\{-\beta E(\{N_\alpha\})\right\}
D(\{N_\alpha\})\prod_\alpha{L\varrho(\theta_\alpha)\Delta\theta_\alpha\choose N_\alpha},
\ee
where $\sum_{\left\{\sum N_\alpha=N\right\}}$ means summation over all possible $N_\alpha$ sets consisting of non-negative elements satisfying $\sum N_\alpha=N$. Finally
\be
Z=
\dots\!\!\sum_{N_\alpha\geq0}\!\!\dots\,\,\exp\left\{-\beta E(\{N_\alpha\})\right\}D(\{N_\alpha\})\prod_\alpha{L\varrho(\theta_\alpha)\Delta\theta_\alpha\choose N_\alpha}. 
\ee
We emphasize, that the factor (\ref{thetaintegralresult}) appearing in the partition function is only formally a combinatorial factor, as it is rather the result of some $\theta$ integrals. (For this for example $L\varrho(\theta_\alpha)\Delta\theta_\alpha$ need not to bee an integer.)

\item
At this point we modify the notations to make them coherent with those of the earlier works \cite{Woyna}: we define the densities according to
\be\label{ro}
N_\alpha=L\rho(\theta_\alpha)\Delta\theta_\alpha,
\ee
and introduce the density of holes according to
\be
\varrho(\theta_\alpha)=\rho(\theta_\alpha)+\rho_h(\theta_\alpha).
\ee
With this we arrive at the (partly) familiar expressions
\bea
&&\rho(\theta)+\rho_h(\theta)=\frac{1}{2\pi}\frac{\partial p(\theta)}{\partial\theta}+\sum_{\alpha}K(\theta-\theta_\alpha)\rho(\theta_\alpha)\Delta\theta_\alpha,\\
&&E(\{N_\alpha\})
=L\sum_{\alpha}{e}(\theta_\alpha)\rho(\theta_\alpha)\Delta\theta_\alpha=E[\rho(\theta)],\\
&&D(\{N_\alpha\})
=\det\left\{\delta_{\alpha\beta}-\frac{K(\theta_\alpha-\theta_\beta)\rho(\theta_\beta)\Delta\theta_\beta}{\rho(\theta_\beta)+\rho_h(\theta_\beta)}\right\}=D[\rho(\theta)]
\eea
and
\bea\label{omegaroroh}
&&\!\!\!\!\!{L\varrho(\theta_\alpha)\Delta\theta_\alpha\choose N_\alpha}\!=\!{L(\rho(\theta_\alpha)\!+\!\rho_h(\theta_\alpha))\Delta\theta_\alpha\choose L\rho(\theta_\alpha)\Delta\theta_\alpha}\!=\!
\omega\left(\rho(\theta_\alpha),\rho_h(\theta_\alpha)\right)\!,\\
&&\!\!\!\!\Omega[\rho(\theta)]=\prod_\alpha^{\phantom{N}}\omega\left(\rho(\theta_\alpha),\rho_h(\theta_\alpha)\right).
\eea

Defining the free energy functional in the usual way (see (\ref{F[rho]}))
we arrive at 
\be\label{modifiedZ}
Z=\sum_{\{\rho(\theta_\alpha)\}}e^{\displaystyle{-\beta F[\rho(\theta)]}}D[\rho(\theta)]
\ee
with $\sum_{\{\rho(\theta_\alpha)\}}$ meaning summation over all the $\rho(\theta_\alpha)$ sets obtained from the $\{N_\alpha\}$
sets through (\ref{ro}).
\end{enumerate}

Actually this (\ref{modifiedZ}) is the expression we wanted to derive. It differs from (\ref{Fsum}) (what is practically the starting point of \cite{Woyna}) in the factor of $D[\rho(\theta)]$, 
meaning that contrary to the naive application of Yang's ideas \cite{CNYCPY} (\ref{Fsum}) is able to give the partition function up to a normalization factor only.

From this point the calculation goes as in \cite{Woyna}.
We replace the sum over $\{\rho(\theta_\alpha)\}$ by integrals:
\be
\sum_{\{\rho(\theta_\alpha)\}}\longrightarrow \int\dots\int\prod_\alpha\left(L\Delta\theta_\alpha d\rho(\theta_\alpha)\right),
\ee
thus we have
\be
Z=\int\dots\int e^{\displaystyle{-\beta F[\rho(\theta)]}}D[\rho(\theta)]\prod_\alpha\left(L\Delta\theta_\alpha d\rho(\theta_\alpha)\right).
\ee
Next in the free energy functional $F$ the entropy term $\ln\Omega[\rho(\theta)]$ is expressed by Stirling's formula containing also the terms next to leading order.
The macroscopic part of $F[\rho(\theta)]$ is expanded up to second order in $r(\theta)=\rho(\theta)-\rho_0(\theta)$ around the equilibrium density $\rho_0(\theta)$, and the Gaussian integral obtained this way is evaluated. In this procedure the sub-macroscopic terms of the entropy (that in fact regularize the functional integral) and the determinant $D[\rho(\theta)]$ can be taken at $\rho_0(\theta)$. 
The latter taken out of the integral yields
\be\label{Zredy}
Z=D[\rho_0(\theta)]\int\dots\int e^{\displaystyle{-\beta F[\rho(\theta)]}}\prod_\alpha\left(L\Delta\theta_\alpha d\rho(\theta_\alpha)\right)
\ee
The evaluation of the integral in (\ref{Zredy}) is given in details in \cite{Woyna} and is recited also in \cite{Pozsgay}, thus we do not repeat it here, just cite the result:
\be\label{Zredyindeed}
Z={\cal N}e^{\displaystyle{-\beta F_{min}+\Delta S}}.
\ee 
Here $F_{min}$ is the macroscopic free energy given by (\ref{Fmin}), $\Delta S$ is the correction due to the contributions of the states near to the equilibrium (saddle point fluctuations)
\be
\Delta S=-\ln\det\left\{\delta_{\alpha\beta}-\frac{K(\theta_\alpha-\theta_\beta)\Delta\theta_\beta}{\displaystyle{1+e^{\displaystyle{\beta\epsilon(\theta_\beta)}}}}\right\},
\ee
and
\be
{\cal{N}}=D[\rho_0(\theta)]
\ee
is actually the same as the normalization factor proposed in \cite{Pozsgay}: due to (\ref{densityratio})
\be\label{Drho0}
D[\rho_0(\theta)]=\det\left\{\delta_{\alpha\beta}-\frac{K(\theta_\alpha-\theta_\beta)\Delta\theta_\beta}{\displaystyle{1+e^{\displaystyle{\beta\epsilon(\theta_\beta)}}}}\right\}.
\ee
As a consequence, although the origins of ${\cal N}$ and $e^{\Delta S}$ are completely different, for the PBC they cancel each other indeed.

\section{Integrable boundaries}
\label{sec:IBC}

In case of integrable boundaries the system is described by the BA equations 
\be\label{BAeoe}
2Lp(\theta_i)+\varphi_0(\theta_i)+\varphi_L(\theta_i)+\sum_{j=1}^N\left(\phi(\theta_i-\theta_j)+\phi(\theta_i+\theta_j)\right)-\phi(2\theta_i)=2\pi I_i,
\ee
where the $\varphi_{0/L}(\theta)$ are phase shifts due to the elastic scatterings on the ends at 0 resp.~$L$, and the $I_i$ quantum numbers are \textsl{always} positive integers. Now only those solutions are physical, in which all (real) $\theta$'s have different modulus, and none of them is zero. Also in this case the energy is given by (\ref{energy}), but now
\be\label{Z_Nbar}
\bar{Z}= \sum_{N=1}^\infty\bar{Z}_N\quad\mbox{with}\quad\bar{Z}_N=\!\!\!\!\!\!\!\sum_{1\leq I_1<I_2<\dots<I_N}\!\!\!\!\!\!\!\exp{\left\{-\beta E(\theta_1,\theta_2,\dots,\theta_N)\right\}}.
\ee 

In calculating the partition function we follow the same program as for the PB case, but some steps of the calculation have to be modified, as this case differs from that of the PBC in two points, both affecting the $O(1)$ corrections. One of these is the different structure of the BAE, the other is that the summation is restricted to the positive integers only. 

Due to the modified structure of the BAE a different determinant will appear in the partition function. When allowing for the quantum numbers being equal and treating them as continuous variables the equation connecting the $I_j$ and $\theta_i$ variables are 
\be\label{mmBAe}
2Lp(\theta_i)+\varphi_0(\theta_i)+\varphi_L(\theta_i)+\sum_{j}^n\left(\phi(\theta_i-\theta_j)+\phi(\theta_i+\theta_j)\right)p_j-\phi(2\theta_i)=2\pi I_i.
\ee
This gives a Jacobi determinant  
\be\label{Jacobi2}
   \det\left\{\frac{\partial I_i}{\partial\theta_j}\right\}=\det\{L\bar{\varrho}(\theta_i)\delta_{ij}-K^-(\theta_i,\theta_j)p_j\}
   \ee
with
\be
\bar{\varrho}(\theta_i)=\sigma(\theta_i)
+\frac{1}{L}\sum_{j=1}^nK^+(\theta_i,\theta_j)p_j,
\ee
where
\be
\sigma(\theta)=\frac{1}{\pi}\frac{\partial p(\theta)}{\partial\theta}
+\frac{1}{2\pi L}\frac{\partial\varphi_0(\theta)}{\partial\theta}
+\frac{1}{2\pi L}\frac{\partial\varphi_L(\theta)}{\partial\theta}-
\frac{2}{L}K(2\theta),
\ee
and
\be
K^{\pm}(\theta,\theta')=K(\theta-\theta')\pm K(\theta+\theta').
\ee
Factorizing it in analogy with (\ref{factorPBC}) leads to 
\be
\bar{D}^{(P)}(\theta_1,\theta_2,\dots\theta_{n_P})\prod_{j=1}^{n_P} L\bar{\varrho}(\theta_j),
\ee
with
\be        
\bar{D}^{(P)}(\theta_1,\dots\theta_n)=
\det\left\{\delta_{ij}-\frac{K^-(\theta_i,\theta_j)p_j}{L\bar{\varrho}(\theta_j)}\right\}.
\ee
This appears in the final expression for $\bar{Z}$ in the form
\be
\bar{D}[\bar{\rho}(\theta)]=\det\left\{\delta_{\alpha\beta}-\frac{K^-(\theta_\alpha,\theta_\beta)\bar{\rho}(\theta_\beta)\Delta\theta_\beta}{\bar{\rho}(\theta_\beta)+\bar{\rho_h}(\theta_\beta)}\right\}
\ee
with
\be
\bar{\rho}(\theta)+\bar{\rho}_h(\theta)=\sigma(\theta)
+\sum_{\alpha}K^+(\theta,\theta_\alpha)\rho(\theta_\alpha)\Delta\theta_\alpha.
\ee

The restriction $0<I_1<I_2\dots$ affects the combinatorial factor entering into the free energy functional, and the careful inspection of it is needed to have the "regular" $O(1)$ terms correctly. Due to this restriction the proper representation of the sums by integrals reads (see Appendix \ref{sec:IB}): 
	\be
\sum_{I_1=1}\sum_{I_2=1}\dots\sum_{I_n=1}\Longrightarrow\int\limits_0\int\limits_0\dots\int\limits_0\prod_{i=1}^n\left(1-\delta(I_i)\right)dI_i,
  \ee
with $\delta(I)$ being the Dirac $\delta$-function, and it is understood, that its integral on the limit of the integration interval is 1/2. Changing the integration variables from the $I$'s to the $\theta$'s, and taking into account that
$\delta(I)=\delta(\theta) /(L\bar{\varrho}(\theta))$ we get
\bea
&&\int\limits_0\int\limits_0\dots\int\limits_0\prod_{i=1}^{n_P}\left(1-\delta(I_i)\right)dI_i\Longrightarrow\nn\\
&&\int\limits_0\int\limits_0\dots\int\limits_0
\left(\prod_{j=1}^{n_P} \left(L\bar{\varrho}(\theta_j)-\delta(\theta_j)\right)d\theta_j\right)\bar{D}^{(P)}(\theta_1,\theta_2,\dots\theta_{n_P}).
\eea
Completing the integrals by the $\Delta\theta_\alpha$ intervals we end up at the combinatorial factor 
\be
\bar{\Omega}[\bar{\rho}(\theta)]=
{L(\bar{\rho}(\theta_0)+\bar{\rho}_h(\theta_0))\Delta\theta_0-1/2\choose L\bar{\rho}(\theta_0)\Delta\theta_0}
\prod_{\alpha>0}^{\vphantom{N}}\omega\left(\bar\rho(\theta_\alpha),\bar{\rho}_h(\theta_\alpha)\right)
\ee
with $\omega(\rho,\rho_h)$ given by (\ref{omegaroroh}) and the $\alpha=0$ index referring to the $\Delta\theta$ interval starting at the origin. With this the free energy reads as 
\be
\bar{F}[\bar{\rho}(\theta)]=E[\bar{\rho}(\theta)]-T\ln\bar{\Omega}[\bar{\rho}(\theta)].
\ee

As a result the partition function is of the form 
\be\label{modifiedZbar}
\bar{Z}=\sum_{\{\bar{\rho}(\theta_\alpha)\}}e^{\displaystyle{-\beta\bar{F}[\bar{\rho}(\theta)]}}\bar{D}[\bar{\rho}(\theta)],
\ee
that is in complete analogy with (\ref{modifiedZ}) and can be treated the same way
leading to
\be\label{Zredyindeedbar}
\bar{Z}=\bar{\cal N}e^{\displaystyle{-\beta \bar{F}_{min}+\overline{\Delta S}}}.
\ee 
Here $\bar{F}_{min}$ is the bulk free energy modified by the "regular" $O(1)$ corrections due to the reflections on the boundaries and the exclusion of the zero rapidity:
\be
 \bar{F}_{min}={F}_{min}+\Delta F + \phi_0+\phi_L
\ee
with 
\be
\Delta F=\frac{T}{2}\ln\left(1+e^{\displaystyle{-\beta\epsilon(0)}}\right)+2T\int\limits_o^\infty K(2\theta)\ln\left(1+e^{\displaystyle{-\beta\epsilon(\theta)}}\right)d\theta,
\ee
and
\be
\phi_{0/L}=-\frac{T}{2\pi}\int\limits_o^\infty \frac{\partial\varphi_{0/L}(\theta)}{\partial\theta}\ln\left(1+e^{\displaystyle{-\beta\epsilon(\theta)}}\right)d\theta    ;
\ee
the $\overline{\Delta S}$ correction due to the saddle point fluctuations is given by
\be
e^{\displaystyle{\overline{\Delta S}}}=\left(\det\left\{\delta_{\alpha\beta}-\frac{K^+(\theta_\alpha,\theta_\beta)\Delta\theta_\beta}{\displaystyle{1+e^{\displaystyle{\beta\epsilon(\theta_\beta)}}}}\right\}\right)^{-1};
\ee
and (as (\ref{densityratio}) with the same $\epsilon$ holds for $\bar{\rho}_0(\theta)$ and $\bar{\rho}_{h,0}(\theta)$ too) the factor $\bar{\cal N}$ is 
\be\label{Dbarrho0}
\bar{\cal N}
=\bar{D}[\bar{\rho}_0(\theta)]=\det\left\{\delta_{\alpha\beta}-\frac{K^-(\theta_\alpha,\theta_\beta)\Delta\theta_\beta}{\displaystyle{1+e^{\displaystyle{\beta\epsilon(\theta_\beta)}}}}\right\}.
\ee

In this case the normalization factor $\bar{\cal N}$ and the contribution of the saddle point fluctuations $e^{\overline{\Delta S}}$ do not cancel each other, but after a straightforward manipulation lead to the expected boundary independent correction
\bea
&&\overline{\Delta S}+\ln\bar{\cal N}=\nn\\
&&\sum_{n=1}^\infty\frac{1}{n}\int\limits_{-\infty}^{\infty}\dots\int\limits_{-\infty}^{\infty}\prod_{i=1}^n\left(\frac{d\theta_i}{{\displaystyle{1+e^{\beta\epsilon(\theta_i)}}}}\right)K(\theta_1+\theta_2)K(\theta_2-\theta_3)\dots K(\theta_n-\theta_1)
\eea
indeed.

\section{Concluding remarks}

In the present work we revisited the calculation of the partition function for certain BA solvable models at periodic boundary conditions and also at integrable reflective boundaries. Our motivation for this was, that the earlier works starting from the free energy functional given in terms of rapidity densities, and based on the calculation of the saddle point fluctuations \cite{Woyna} turned out to miss important $O(1)$ corrections to the free energies of these systems \cite{Pozsgay}. To derive the missing pieces we started by the exact expression of the partition function given as a sum over the possible quantum numbers entering into the BA equations ((\ref{Z_N}) and (\ref{Z_Nbar})). In several steps (representing these sums by integrals over the variables obtained by considering the quantum numbers as continuous variables, changing the integration variables to the rapidities, and completing this integrals in a special way) we transformed this expression into a one given as a sum over the possible rapidity densities. This way we have shown, that the correct expression for the partition function is the one given in terms of the rapidity densities in the usual way, but completed by a density dependent normalization factor arising due to the Jacobi determinant connected to the change of integration variables ((\ref{modifiedZ}),(\ref{modifiedZbar})). This factor (being $O(1)$ in itself) can be taken at the equilibrium density to yield the nontrivial normalization factor proposed intuitively in \cite{Pozsgay} ((\ref{Drho0}),(\ref{Dbarrho0})).

An interesting feature of the calculation is that the combinatorial factors giving the entropy part of the free energy functional are obtained as the results of some integrals over the rapidities themselves. As these integrals are slightly different for the PBC and the IB case, we have presented the calculation for both cases. The main point, however, the appearance of the nontrivial normalization has the same origin, namely the change of the variables from the quantum numbers to the rapidities, and inspecting the slight differences in the integrals are needed not to miss the "normal" $O(1)$ corrections in the IB case.

Since the fundamental work of C.N.~Yang and C.P.~Yang \cite{CNYCPY} most of the thermodynamic descriptions of Bethe Ansatz systems have been formulated in terms of the rapidity densities. A different approach of the problem was presented by Kato and Wadati \cite{KaWa}, who calculated the $N$-particle cluster integrals based on the partition function (\ref{Z}-\ref{Z_N}), and have found complete agreement with the results of thermodynamic Bethe Ansatz. It seems, however, that such an equivalence exists on the macroscopic level only, and the thermodynamic Bethe Ansatz formulated in the usual way is not accurate enough to give the sub macroscopic contributions correctly. To be definite, the combinatorial factor (\ref{Omegarho})
\be\label{Omegarho2} 
\Omega[\rho(\theta)]=\prod{L(\rho(\theta)+\rho_h(\theta))\Delta\theta\choose L\rho(\theta)\Delta\theta}
\ee
is able to give the number of states represented by a rapidity density $\rho(\theta)$ in leading order only, and the expression giving also the next to leading order correctly is of the form   
\be\label{numberofstates} 
D[\rho(\theta)]\,\Omega[\rho(\theta)]\!=\!\det\left\{\delta_{\alpha\beta}-\frac{K(\theta_\alpha,\theta_\beta)\rho(\theta_\beta)\Delta\theta_\beta}{\rho(\theta_\alpha)\!+\!\rho_h(\theta_\alpha)}\right\}\prod{L(\rho(\theta)\!+\!\rho_h(\theta))\Delta\theta\choose L\rho(\theta)\Delta\theta}.
\ee
Here the appearance of $D[\rho(\theta)]$ results in the normalization factor proposed in \cite{Pozsgay}.

\section*{Acknowledgments}
I am grateful to Dr.~Z.~Bajnok for the critical reading of the manuscript. The support of OTKA under grant Nr.~K68340 is acknowledged.

\appendix
\section{}\label{sec:summation}
Our aim is to find an accurate enough integral representation of a sum of the type $\sum_{I_1<I_2<\dots<I_N}f(I_1,I_2,\dots I_N)$ with symmetric summand $f(\{I_i\})$ and $I_j$'s being integers or half-odd-integers. The replacement 
	\be\label{summa}
	\sum_{I_1<I_2<\dots<I_N}f(I_1,I_2,\dots I_N)\Longrightarrow\int_{I_1<I_2<\dots<I_N}f(I_1,I_2,\dots I_N)\prod_{i=1}^NdI_i
\ee
is obviously too rough for our purposes: if we suppose, that the $I_j$ are the half-odd-integers between zero and some integer $M$, and we take $f(\{I_i\})\equiv1$, then the value of the left-hand side is $M\choose N$, while the right-hand side (with limits zero and $M$) gives $M^N/N!$. The difference is caused by the improper treatment of the exclusions $I_1\not=I_2$, $I_2\not=I_3,\dots I_{N-1}\not=I_N$.
Now we present a systematic way to solve this problem. As the treatment of an $I_i<I_j$ restriction in an integral gives different result as in a sum, first we represent the left-hand side of (\ref{summa}) in a form, which does not contain such restrictions, and in a second step we analyze the properties of the integral obtainable this way.

We start with the sum
\be
\Sigma_0=\frac{1}{N!}\sum_{I_1}\sum_{I_2}\cdots\sum_{I_N}f(I_1,I_2,\dots I_N)
\ee
in which the contribution of the configurations with all $I_j$ being different is the same as required, but it contains also the contribution of the terms with some of the $I_j$'s equal. For this we subtract a sum in which the the configurations with different $I$'s do not contribute, but it eliminates at least the unwanted $I_i=I_j$ contributions:
\be
\Sigma_1=-\frac{1}{N!}{N\choose2}\sum_{I_1}\sum_{I_2}\cdots\sum_{I_{N-1}}f(\underbrace{I_1,I_1,}I_2,\dots I_{N-1}).
\ee
In $\Sigma_0+\Sigma_1$ the configurations with exactly two $I$'s equal do not contribute, but it still counts the contributions of the configurations with more than two $I$'s equal. Next we eliminate from $\Sigma_0+\Sigma_1$ the contributions of the configurations of the types $I_i=I_j$, $I_k=I_l$ and $I_i=I_j=I_k$ by adding: 
\bea
\Sigma_2&=&\frac{1}{N!}\frac{1}{2}{N\choose2}{N-2\choose2}\sum_{I_1}\sum_{I_2}\sum_{I_3}\cdots\sum_{I_{N-2}}f(\underbrace{I_1,I_1,}\underbrace{I_2,I_2,}I_3,\dots I_{N-2})\nn\\
&+&2\frac{1}{N!}{N\choose3}\sum_{I_1}\sum_{I_2}\cdots\sum_{I_{N-2}}f(\underbrace{I_1,I_1,I_1,}I_2,\dots I_{N-2}).
\eea
In $\Sigma_0+\Sigma_1+\Sigma_2$ none of the configurations, which are described by one or two equality do not contribute. The configurations described by three equalities are of the type i): $I_i=I_j$, $I_k=I_l$ and $I_m=I_n$, ii): $I_i=I_j=I_k$ and $I_l=I_m$, or iii): $I_i=I_j=I_k=I_l$. These can be eliminated by adding
\bea
\Sigma_3&=&-\frac{1}{N!}\frac{1}{3!}{N\choose2}{N-2\choose2}{N-4\choose2}\sum_{I_1}\sum_{I_2}\sum_{I_3}\sum_{I_4}\cdots\nn\\
&&\cdots\sum_{I_{N-3}}f(\underbrace{I_1,I_1,}\underbrace{I_2,I_2,}\underbrace{I_3,I_3,}I_4,\dots I_{N-3})\nn\\ 
&-&2\frac{1}{N!}{N\choose3}{N-3\choose2} \sum_{I_1}\sum_{I_2}\sum_{I_3}\cdots\sum_{I_{N-3}}f(\underbrace{I_1,I_1,I_1,}\underbrace{I_2,I_2,}I_3,\dots I_{N-3})\nn\\
&-&6\frac{1}{N!}{N\choose4}\sum_{I_1}\sum_{I_2}\cdots\sum_{I_{N-3}}f(\underbrace{I_1,I_1,I_1,I_1,}I_2,\dots I_{N-3}).
\eea
Now the systematics is clear: in each step we eliminate the contributions with the minimum number of equalities. For example in
\be
\sum_{i=0}^n \Sigma_i
\ee
the contribution of the configurations with all $I$'s different is as required, but it contains unwanted contributions from configurations in which groups of $I$'s containing all together more than $n+1$ elements are equal. The $\Sigma_{n+1}$ is chosen to eliminate the contributions of those configurations, which can be characterized by $n+1$ equality signs "=". The procedure is in each step combinatorially well defined, and it leads to the form 
\be\label{osszA}
	 \sum_{I_1<I_2<\dots<I_N}f(I_1,I_2,\dots I_N)=\sum_P a(P)\sum_{I_1,I_2,\dots,I_{n_P}}f^{(P)}(I_1,I_2,\dots I_{n_P}).
	\ee
	Here the summation on the right-hand side (r.h.s.) goes over the partitions $P$ of the number $N$, the $a(P)$ coefficients are constructed by the above procedure, and in the functions $f^{(P)}$ groups of the $I_i$ parameters corresponding to $P$ are equal: for example for a partition $P:\{p_1+p_2+\dots+p_n\}$ 	
  \be\label{ossz2A}
  f^{(P)}(I_1,I_2,\dots I_n)=f(\overbrace{\underbrace{I_1,\dots, I_1}_{p_1},\underbrace{I_2,\dots,I_2}_{p_2},\dots,\underbrace{I_n,\dots,I_n}_{p_n}}^N).
  \ee

We note here, that the (\ref{osszA}) representation of the sums over the quantum numbers is closely related to the one introduced in \cite{KaWa},
but it is formulated in a less abstract way. Comparing the two yields an expression for the $a(P)$ coefficients:
\be
a(P)=\frac{1}{N!}F(P)C(P)\ ,
\ee
where 
\be
F(P)=\prod_{i=1}^n(-1)^{(p_i-1)}(p_i-1)!\,,
\ee
as given in \cite{KaWa}, and $C(P)$ is the combinatorial factor giving the number of ways the sets of $p_1$, $p_2$, \dots $p_n$ elements can be chosen out of the $N$ ones irrespective of their order. If in the partition  $P:\{p_1+p_2+\dots+p_n\}$ the element $p_j$ is present $\nu_j$ times, i.e. $\sum_j \nu_jp_j=N$ and $\sum_j \nu_j=n$, then 
\be
C(P)=N!\prod_j\frac{1}{\nu_j!(p_j!)^{\nu_j}}\ ,
\ee
and
\be\label{a(P)general}
a(P)=\prod_{j}(-1)^{\nu_j(p_j-1)}\frac{1}{\nu_j!{p_j}^{\nu_j}}\ .
\ee
In our work we do not use the explicit form of $a(P)$, only the fact, that such a representation exists, and some properties of this representation discussed below are exploited.

\section{}\label{sec:partialsum}
An important property of the (\ref{osszA}) representation can be obtained by applying a special function for $f(\{I\})$:
\be
f(I_1,\dots,I_N)=\left(\prod_{i=1}^Nh(I_i)\right)g(I_1,\dots,I_N)\quad\mbox{with}\quad h(I)=\left\{\ba{ll}1,&\mbox{if $I<M$,}\\
x,&\mbox{if $I>M$,}\ea\right.
\ee 
where $x$ is an auxiliary variable and $M$ is a suitable number between the lower and upper limits of summations, not equal to any of the possible $I$'s. Substituting this both sides of (\ref{osszA}) become polynomials of the variable $x$:
\bea
&&\sum_{i=0}^N\left(\sum_{I_1<I_2<\dots<I_{i}<M}\ \sum_{M<I_{i+1}<I_{i+2}<\dots<I_{N}}x^{N-i}g(I_1,I_2,\dots I_N)\right)=\nn\\
&&\sum_P a(P)\left(\sum_{I_1<M}+x^{p_1}\!\!\!\!\sum_{M<I_1}\right)\dots\left(\sum_{I_{n_P}<M}+x^{p_{n_P}}\!\!\!\!\sum_{M<I_{n_P}}\right)g^{(P)}(I_1,I_2,\dots I_{n_P}).
\eea
Equating the coefficients of the same $x$ powers we obtain relations of the structure
\bea\label{restricted1}
&&\sum_{I_1<\dots<I_{N_1}<M}\ \sum_{M<I_{N_1+1}<\dots<I_{N}}g(I_1,I_2,\dots I_N)=\sum_{P=P_1\oplus P_2} a(P)\,c(P_1\,,P_2)\times\nn\\
&&\sum_{I_1,\dots,I_{n_{P_1}}<M}
\sum_{M<I_{n_{P_1}+1},\dots,I_{n_{P_1}+n_{P_2}}}g^{(P_1\oplus P_2)}(I_1,I_2,\dots I_{n_{P_1}+n_{P_2}}). 
\eea
Here $N_1$ is an integer less than $N$, $P_1$ and $P_2$ are partitions of the numbers $N_1$ and $N_2=N-N_1$, respectively; $P=P_1\oplus P_2$ is the partition of $N$ emerging as a composition of $P_1$ and $P_2$, and $\sum_{P=P_1\oplus P_2}$ means summation over all of these partitions; $c(P_1\,,P_2)$ is a combinatorial factor giving the number of ways $P$ can be split up into parts $P_1$ and $P_2$; finally $g^{(P_1\oplus P_2)}(I_1,I_2,\dots I_{n_{P_1}+n_{P_2}})$ is the same as $g^P$, just the variables are permuted to have the ones corresponding to $P_1$
(those which should be less than $M$) appearing at the first $n_{P_1}$ positions. This shows, that if on the l.h.s.~of (\ref{osszA}) a restriction like in (\ref{restricted1}) is imposed, only those terms of the r.h.s.~contribute which are compatible with this restriction in the above sense. This is, however, true on the opposite way around too: if on the r.h.s.~of (\ref{osszA}) only those terms are taken into account in which the variables $I_j$ counted with the multiplicities $p_j$ can be grouped into sets of elements $N_1$ resp.~$N_2=N-N_1$, and the summation over the variables belonging to different sets is carried out over nonoverlapping regions like in the r.h.s.~of (\ref{restricted1}), then the result will correspond to a partial sum like that on the l.h.s.~of (\ref{restricted1}). 

(\ref{osszA}) can be applied to the $I_1<\dots<I_{N_l}<M$ and $M<I_{N_l+1}<\dots<I_N$ sums on the l.h.s.~of (\ref{restricted1}). This leads to the relation 
\be
a(P_1)a(P_2)=a(P_1\oplus P_2)c(P_1,P_2),
\ee
that can be checked directly for the coefficients given explicitly in Appendix \ref{sec:summation}, and it holds also for the general formula (\ref{a(P)general})

\section{}\label{sec:integral}
Now we replace in (\ref{osszA}) the summations by integrals
\be\label{substitution}
\sum_{I_1<I_2<\dots<I_N}f(I_1,I_2,\dots I_N)\Longrightarrow\sum_P a(P)\int f^{(P)}(I_1,I_2,\dots,I_{n_P})\prod_{i=1}^{n_P} dI_i\,.
\ee
From the fact, that for certain cases the above replacement is exact, some important properties of the integral of the right-hand side can be derived. 

Suppose, that the $I_j$ numbers are half odd-integers between zero and some integer $M$, and $f(I_1,I_2,\dots,I_N)\equiv1$.
The substitution (\ref{substitution}) is exact, if the integral limits are $0$ and $M$. From this we have 
\be\label{suma(P)}
\sum_P a(P) M^{{\displaystyle{n}}_P}={M\choose N}.
\ee
As, however, the structure of the polynomial on the l.h.s.~is independent of the integration limits,
\be
\sum_P a(P)\int_x^y f^{(P)}(I_1,I_2,\dots,I_{n_P})\prod_{i=1}^{n_P} dI_i=\sum_P a(P) L^{{\displaystyle{n}}_P}={L\choose N}
\ee
for any $y-x=L$.

In a similar way, we may require that in the l.h.s.~of (\ref{substitution}) the first $N_1$ of the $I_j$'s should be less than the integer $M_1$, and the rest of them fall between $M_1$ and $M$ (like in Appendix \ref{sec:partialsum}). In this case only those partitions on the r.h.s. contribute, which can be split up into partitions of $N_1$ and $N_2=N-N_1$ (i.e.~$P=P_1\oplus P_2$ with $P_1$ and $P_2$ being partitions of $N_1$ resp.~$N_2$). The integration limits should be taken as $0$ and $M_1$ for the variables belonging to $P_1$, and they should be $M_1$ and $M$ for the others. The result of the integration is a polynomial of $M_1$ and $M_2=M-M_1$ which now is 
\be
{M_1\choose N_1}{M_2\choose N_2}
\ee
It follows from this, that if we take the integral such a way, that $N_1$ variable should fall into an interval of length $L_1$ and the rest into an other nonoverlapping (or even disjoint) one with length $L_2$, then the result is 
\be
{L_1\choose N_1}{L_2\choose N_2}
\ee
This property can be generalized to any number of nonoverlapping intervals: if the integration is carried out under the restriction,  that $N_\alpha$ of the variables (counted with the proper multiplicity) fall into the interval $L_\alpha$, $\alpha=1,2,\dots$, then the result is
\be\label{integralbyintervals}
\prod_\alpha{L_\alpha\choose N_\alpha}.
\ee

\section{}\label{sec:IB}
In case of the integrable boundaries we have to calculate sums over the $I_i$ quantum numbers taking integer values with restrictions $\{0<I_1, I_i<I_{i+1}\}$. The restriction of the type $I_i<I_{i+1}$ can be dissolved as in the general case leading to
\be\label{osszC}
	 \sum_{0<I_1<I_2<\dots<I_N}f(I_1,I_2,\dots I_N)=\sum_P a(P)\sum_{0<I_1}\sum_{0<I_2}\dots\!\!\!\sum_{0<I_{n_P}}f^{(P)}(I_1,I_2,\dots I_{n_P}),
	\ee
with $P$ being the partitions of $N$. If $M$ is an integer,
\be
\sum_{I=1}^M f(\dots I\dots)\approx \int\limits_{1/2}^{M+1/2}f(\dots I\dots)dI.
\ee
This formula, however, is not convenient for us, as changing the integration variables to the rapidities $\theta$, the limit of integration replacing the $1/2$ will be a function of the other rapidities. For this we use an other formula 
\be
\sum_{I=1}^M f(\dots I\dots)\approx \!\!\!\!\!\int\limits_{0}^{M+1/2}\!\!\!\!\!f(\dots I\dots)dI-\frac{1}{2}f(\dots I=0\dots),
\ee
which, by taking into account, that the integral of the Dirac $\delta$-function on the end of the integration interval is 1/2, reads
\be
\sum_{I=1}^M f(\dots I\dots)\approx \!\!\!\!\!\int\limits_{0}^{M+1/2}\!\!\!\!\!f(\dots I\dots)(1-\delta(I))dI.
\ee
This way we have
\bea\label{intC}
	&& \sum_{0<I_1<I_2<\dots<I_N}^Mf(I_1,I_2,\dots I_N)=\nn\\
	&& \sum_P a(P)\!\!\int\limits_{0}^{M+1/2}\!\!\dots\!\!\int\limits_{0}^{M+1/2}\!\!f^{(P)}(I_1,I_2,\dots I_{n_P})\prod_{i=1}^{n_P}\left(1-\delta(I_i)\right)dI_i.
\eea
For a constant $f(\dots I\dots)$ the integrals on the r.h.s.~can be completed, and due to (\ref{suma(P)}) they yield
\be
\sum_P a(P)\int\limits_{0}^{L}\dots\int\limits_{0}^{L}\prod_{i=1}^{n_P}\left(1-\delta(I_i)\right)dI_i={L-1/2\choose N}
\ee
for any upper limit $L$.

\section{}\label{sec:integrationerror}

The quantitative estimation of the error introduced by the replacement (\ref{int_repr})
  \be\label{int_repr_m}
	\sum_{I_1}\sum_{I_2}\dots\sum_{I_n}f^P(I_1,I_2\dots I_n)\Longrightarrow\int\int\dots\int f^P(I_1,I_2\dots I_n)\prod_{i=1}^n dI_i,
  \ee
may need different considerations for different models. In this Appendix we deal with the $\delta$ Bose gas with PBC, for which the rapidity variables are the wavenumbers (momenta) of the particles ($\theta_i=k_i$) determined by the equations (see (\ref{mBAe}))
\be\label{mBAe_m}
Lk_i+\sum_{j}^n2\tan^{-1}\left(\frac{k_i-k_j}{c}\right)p_j=2\pi I_i,
\ee
the energy is quadratic, i.e.:  
\be
e(k_i)=k_i^2-\mu
\ee  
(with $\mu$ being the chemical potential), thus the $f^P$ function is 
\be
f^P(I_1,I_2\dots I_n)=\exp\left\{-\beta\sum_{j=1}^n p_j\left(k_j^2-\mu\right)\right\}\,.
\ee
We use the Poisson summation formula (see also \cite{KaWa})
\be\label{Psf}
\sum_{I=integer}g(I)=\sum_{J=integer}\int g(I)e^{2\pi iIJ}dI.
\ee
According to this, the (\ref{int_repr}) (or (\ref{int_repr_m})) type representation of the infinite sum is equivalent to taking on the right-hand side the $J=0$ term only, and the error introduced can be estimated by the neglected $J\not=0$ contributions. For each $I$ we apply (\ref{Psf}) (for the sake of simplicity we suppose, that the $I_j$'s are integers), and we calculate a general term 
\be
\int\int\dots\int\exp\left\{-\beta\sum_{j=1}^n p_j\left(k_j^2-\mu\right)+2\pi i\sum_{j=1}^nI_jJ_j\right\}\prod_{j=1}^n dI_j.
\ee
This, after changing the integration variables from the $I$'s to the $k$'s, due to the $\tan^{-1}$ form of the phases reads
\bea  
&&\int\int\dots\int\exp\left\{-\beta\sum_{j=1}^n p_j\left(k_j^2-\mu-iLk_jJ_j/\beta p_j\right)\right\}\times\nn\\
&&\prod_{j,l=1}^n\left(c+i(k_j-k_l)\right)^{J_jp_l-J_lp_j}
\det\left\{\frac{\partial I_j}{\partial k_l}\right\}\prod_l dk_l.
\eea
Deforming the integral contour like $k_j\longrightarrow k_j+i\kappa_j$ with $\kappa_j=LJ_j/2\beta p_j$ we obtain
\bea\label{genterm} 
&&\!\!\exp\left\{-\sum_{j=1}^n \frac{L^2J_j^2}{4\beta p_j}\right\}\int\int\dots\int\exp\left\{-\beta\sum_{j=1}^n p_j\left(k_j^2-\mu\right)\right\}\times\nn\\
&&\!\!\prod_{j,l=1}^n\!\left(c\!-\!(\kappa_j\!-\!\kappa_l)\!+\!i(k_j\!-\!k_l)\right)^{J_jp_l-J_lp_j}
\left.\!\det\left\{\frac{\partial I_j}{\partial k_l}\right\}\right\vert_{k_l\to k_l+i\kappa_l}\prod_l dk_l.
\eea
It is important, that with the contour deformation no pole contribution is collected. (To see this we need the following observations. The first is, that in the different terms of $\left.\det\left\{{\partial I_j}/{\partial k_l}\right\}\right\vert_{k_l\to k_l+i\kappa_l}$ each   $\left(c-(\kappa_j-\kappa_l)+i(k_j-k_l)\right)$ has a power zero or $-1$, those terms with $\left(c-(\kappa_j-\kappa_l)+i(k_j-k_l)\right)^{-2}$ cancel each other. The second is that if $(\kappa_j-\kappa_l)>0$ (thus $c-(\kappa_j-\kappa_l)$ can be negative), then ${J_jp_l-J_lp_j}\geq1$. The result is, that the contour deformation can be done such a way, that none of the zeros of the denominators are crossed.) Due to the prefactor in (\ref{genterm}) the order of magnitude of a general  $J_1,J_2,\dots J_n$ term is $\exp\left\{-\sum_{j=1}^n {L^2J_j^2}/{4\beta p_j}\right\}$ times smaller, than that of the leading all $J=0$ term, i.e.~supposing that $p_1$ is the largest element in the partition $P$, even the largest correction to (\ref{int_repr_m}) means an 
\be
O\left(\exp\left\{-\frac{L^2}{4\beta p_1}\right\}\right)
\ee
relative error. As the possible maximum of $p_1$ is $N$, this is well (over)estimated by the partition independent expression
\be\label{overestimation}
O\left(\exp\left\{-\frac{L^2}{4\beta N}\right\}\right).
\ee
For those terms, which contribute to the partition function $N/L$ is finite as $N,L\to\infty$, i.e.~(\ref{overestimation}) decays exponentially with $L$, thus we may conclude, that the integral representation of the sums over the quantum numbers is a very good approximation for the $\delta$ Bose gas.

\end{document}